\documentclass[a4paper, 10 pt, conference]{ieeeconf}  % Comment this line out
\IEEEoverridecommandlockouts                              % This command is only
\usepackage{graphicx,amssymb,amstext}
\usepackage{amsmath}
\usepackage{hyperref}
\usepackage{epsfig}
\usepackage{verbatim}
\usepackage{cite}

\newtheorem{theorem}{Theorem}%[section]
\title{\LARGE \bf
Robust $H_\infty$ Coherent-Classical Estimation of Linear Quantum Systems
}

\author{Shibdas Roy* and Ian R. Petersen%
\thanks{This work was supported by the Australian Research Council and the Singapore National Research Foundation.}%
\thanks{S.~Roy is with the Department of Electrical and Computer Engineering, National University of Singapore, Singapore, and I.~R.~Petersen is with the School of Engineering and Information Technology, University of New South Wales, Canberra, Australia.}%
\thanks{*\tt\small roy\_shibdas at yahoo.co.in}%
}

\begin{document}

\maketitle
\thispagestyle{empty}
\pagestyle{empty}

%%%%%%%%%%%%%%%%%%%%%%%%%%%%%%%%%%%%%%%%%%%%%%%%%%%%%%%%%%%%%%%%%%%%%%%%%%%%%%%%
\begin{abstract}

We study robust $H_\infty$ coherent-classical estimation for a class of physically realizable linear quantum systems with parameter uncertainties. Such a robust coherent-classical estimator, with or without coherent feedback, can yield better disturbance-to-error performance than the corresponding robust purely-classical estimator for an uncertain plant. Moreover, coherent feedback allows for such a robust coherent-classical estimator to be more robust to uncertainty in comparison to the robust classical-only estimator.

\end{abstract}

%%%%%%%%%%%%%%%%%%%%%%%%%%%%%%%%%%%%%%%%%%%%%%%%%%%%%%%%%%%%%%%%%%%%%%%%%%%%%%%%
\section{Introduction}

\bstctlcite{BSTcontrol}

It has been of significant interest recently to study estimation and control problems for quantum systems \cite{WM1,NY,JNP,NJP,SL}. Linear quantum systems \cite{WM1,NY,JNP,NJP,WM2,GZ} are an important class of quantum systems, and allow for describing quantum optical devices such as finite bandwidth squeezers \cite{GZ}, optical cavities \cite{WM2}, and linear quantum amplifiers \cite{GZ}. Coherent feedback control for linear quantum systems, where the feedback controller is also a quantum system, has been more recently studied \cite{JNP,NJP,SL}. A related coherent-classical estimation problem has been considered by the authors in \cite{IRP2,RPH,RPH1}, where the estimator consists of a classical part, which produces the desired final estimate and a quantum part, which may also involve coherent feedback. A quantum observer, as constructed in \cite{NY}, is a purely quantum system, that produces a quantum estimate of a variable for the quantum plant. In contrast, a coherent-classical estimator is a mixed quantum-classical system, that produces a classical estimate of a variable for the quantum plant.

The authors have previously studied robust $H_\infty$ classical estimation for an uncertain linear quantum system \cite{RP}. In this paper, we apply and extend such a robust $H_\infty$ estimator to the problem of coherent-classical estimation of an uncertain quantum plant. We note that for a suitable choice of the coherent controller, a robust $H_\infty$ coherent-classical estimator may yield improved disturbance attenuation when compared to the classical-only estimation scheme of \cite{RP}. Furthermore, we observe that with the addition of coherent feedback from the controller to the plant, such a robust $H_\infty$ coherent-classical estimator exhibits superior robustness to uncertainty than the purely-classical robust $H_\infty$ estimator.

\section{Robust Purely-Classical Estimation}
\begin{figure}[!b]
\vspace*{-4mm}
\centering
\includegraphics[width=0.45\textwidth]{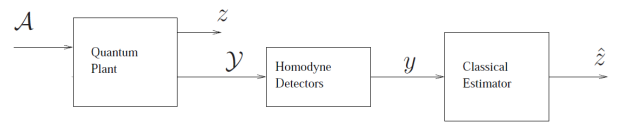}
\caption{Schematic diagram of classical estimation for a quantum plant.}
\label{fig:cls_scm}
%\vspace*{-3mm}
\end{figure}

A schematic diagram of a classical estimation scheme is provided in Fig. \ref{fig:cls_scm}. The quantum plant is defined as linear quantum stochastic differential equations (QSDEs) \cite{RPH}:

\vspace*{-4mm}\small
\begin{equation}\label{eq:plant}
\begin{split}
\left[\begin{array}{c}
da(t)\\
da(t)^{\#}
\end{array}\right] &= A \left[\begin{array}{c}
a(t)\\
a(t)^{\#}
\end{array}\right] dt + B \left[\begin{array}{c}
d\mathcal{A}(t)\\
d\mathcal{A}(t)^{\#} 
\end{array}\right],\\
\left[\begin{array}{c}
d\mathcal{Y}(t)\\
d\mathcal{Y}(t)^{\#}
\end{array}\right] &= C \left[\begin{array}{c}
a(t)\\
a(t)^{\#}
\end{array}\right] dt + D \left[\begin{array}{c}
d\mathcal{A}(t)\\
d\mathcal{A}(t)^{\#}
\end{array}\right],\\
z &= L\left[\begin{array}{c}
a(t)\\
a(t)^{\#} 
\end{array}\right],
\end{split}
\end{equation}
\normalsize\vspace*{-3mm}
where
\vspace*{-3mm}\small
\begin{equation}
\begin{split}
A &= \Omega(A_1,A_2), \qquad B = \Omega(B_1,B_2),\\
C &= \Omega(C_1,C_2), \qquad D = \Omega(D_1,D_2).
\end{split}
\end{equation}
\normalsize\vspace*{-4mm}

Here, $a(t) = [a_1(t) \hdots a_n(t)]^T$ is a vector of annihilation operators. The vector $\mathcal{A}(t) = [\mathcal{A}_1(t) \hdots \mathcal{A}_m(t)]^T$ represents a collection of external independent quantum field operators and the vector $\mathcal{Y}(t)$ represents the corresponding vector of output field operators. Also, $z$ denotes a scalar operator on the underlying Hilbert space and represents the quantity to be estimated. The notation $\Omega(A_1,A_2)$ denotes the matrix $\left[\begin{array}{cc} A_1 & A_2\\
A_2^{\#} & A_1^{\#}
\end{array}\right]$. Moreover, $A_1$, $A_2 \in \mathbb{C}^{n \times n}$, $B_1$, $B_2 \in \mathbb{C}^{n \times m}$, $C_1$, $C_2 \in \mathbb{C}^{m \times n}$, and $D_1$, $D_2 \in \mathbb{C}^{m \times m}$. Furthermore, $^{\#}$ denotes the adjoint of a vector of operators or the complex conjugate of a complex matrix, and $^\dagger$ denotes the adjoint transpose of a vector of operators or the complex conjugate transpose of a complex matrix.

A linear quantum system of the form (\ref{eq:plant}) should satisfy certain physical realizability conditions (See \cite{IRP2,RPH,RPH1}) to represent an actual physical system. A quadrature of each (coherent) component of $\mathcal{Y}(t)$ is measured using homodyne detection to yield a corresponding classical signal $y_i$ \cite{IRP2}:

\vspace*{-4mm}\small
\begin{equation}\label{eq:class_hd}
\begin{split}
dy_1 &= \frac{e^{-\iota\theta_1}}{\sqrt{2}}d\mathcal{Y}_1 + \frac{e^{\iota\theta_1}}{\sqrt{2}}d\mathcal{Y}_1^{*},\\
&\vdots\\
dy_m &= \frac{e^{-\iota\theta_m}}{\sqrt{2}}d\mathcal{Y}_m + \frac{e^{\iota\theta_m}}{\sqrt{2}}d\mathcal{Y}_m^{*}.
\end{split}
\end{equation}
\normalsize\vspace*{-3mm}

Here, $\iota=\sqrt{-1}$, and $\theta_1,\hdots,\theta_m$ determine the quadrature measured by each homodyne detector. The vector of signals $y = [y_1 \hdots y_m]^T$ is then input to a classical estimator.

Corresponding to the system described by (\ref{eq:plant}), (\ref{eq:class_hd}), we define our uncertain system modelled as follows \cite{RP}:

\vspace*{-4mm}\small
\begin{equation}\label{eq:uncertain1}
\begin{split}
\dot{x}(t) &= [A+\Delta A(t)]x(t) + [B+\Delta B(t)]w(t),\\
z(t) &= Lx(t),\\
y'(t) &= S[C+\Delta C(t)]x(t) + SDw(t),
\end{split}
\end{equation}
\normalsize
where $x(t) := \left[\begin{array}{cc}
a(t)^T & a(t)^{\dagger}
\end{array}\right]^T$ is the state, $w(t)$ is the disturbance input, $z(t)$ is a linear combination of the state variables to be estimated, $y'(t)$ is the measured output, $L \in \mathbb{C}^{p \times 2n}$, $SC \in \mathbb{C}^{m \times 2n}$, $SD \in \mathbb{C}^{m \times 2m}$, $S = \left[\begin{array}{cc}
S_1 & S_2
\end{array}\right]$,

\vspace*{-3mm}\small
\begin{equation}
\begin{split}
S_1 &= \left[\begin{array}{cccc}
\frac{e^{-\iota\theta_1}}{\sqrt{2}} & 0 & \hdots & 0\\
0 & \frac{e^{-\iota\theta_2}}{\sqrt{2}} & \hdots & 0\\
 & & \ddots & \\
 & & & \frac{e^{-\iota\theta_m}}{\sqrt{2}}
\end{array}\right],\\
S_2 &= \left[\begin{array}{cccc}
\frac{e^{\iota\theta_1}}{\sqrt{2}} & 0 & \hdots & 0\\
0 & \frac{e^{\iota\theta_2}}{\sqrt{2}} & \hdots & 0\\
 & & \ddots & \\
 & & & \frac{e^{\iota\theta_m}}{\sqrt{2}}
\end{array}\right],
\end{split}
\end{equation}\normalsize
and $\Delta A(\cdot)$, $\Delta B(\cdot)$ and $\Delta C(\cdot)$ denote the time-varying parameter uncertainties, that have the following structure:

\vspace*{-3mm}\small
\begin{equation}\label{eq:unc_pars}
\begin{split}
\left[\begin{array}{c}
\Delta A(t)\\
\Delta C(t)
\end{array}\right] &= \left[\begin{array}{c}
H_1\\
H_3
\end{array}\right]F_1(t)E,\\
\Delta B(t) &= H_2F_2(t)G,
\end{split}
\end{equation}
\normalsize%\vspace*{-3mm}
where $H_1$, $H_2$, $H_3$, $E$ and $G$ are known complex constant matrices with appropriate dimensions, and the unknown matrix functions $F_1(\cdot)$ and $F_2(\cdot)$ satisfy the following bounds:

\vspace*{-1mm}\small
\begin{equation}\label{eq:unc_constraint}
F_1^\dagger(t)F_1(t) \leq I, \, F_2^\dagger(t)F_2(t) \leq I, \qquad \forall t.
\end{equation}
\normalsize

In addition, the uncertainties $\Delta A(t)$, $\Delta B(t)$ and $\Delta C(t)$ should satisfy certain constraints for the uncertain system (\ref{eq:uncertain1}) to be physically realizable (See \cite{RP}).

The robust $H_\infty$ estimation problem for the uncertain system (\ref{eq:uncertain1}) can be converted into a scaled $H_\infty$ control problem, similar to \cite{FDX}, by introducing the following parameterized linear time-invariant system corresponding to (\ref{eq:uncertain1}) \cite{RP}:

\small
\begin{equation}\label{eq:uncertain2}
\begin{split}
\dot{x}(t) &= Ax(t) + \left[\begin{array}{ccc}
B & \frac{\gamma}{\epsilon_1}H_1 & \frac{\gamma}{\epsilon_2}H_2
\end{array}\right]\tilde{w}(t),\\
\tilde{z}(t) &= \left[\begin{array}{c}
\epsilon_1E\\
0\\
L
\end{array}\right]x(t) + \left[\begin{array}{ccc}
0 & 0 & 0\\
\epsilon_2G & 0 & 0\\
0 & 0 & 0
\end{array}\right]\tilde{w}(t)+ \left[\begin{array}{c}
0\\
0\\
-I
\end{array}\right]u(t),\\
y'(t) &= SCx(t) + \left[\begin{array}{ccc}
SD & \frac{\gamma}{\epsilon_1}SH_3 & 0
\end{array}\right]\tilde{w}(t).
\end{split}
\end{equation}
\normalsize

Here, $u(t)$ is the control input, $\tilde{z}(t)$ is the controlled output, $\epsilon_1$, $\epsilon_2 > 0$ are suitably chosen scaling parameters and $\gamma > 0$ is the desired level of disturbance attenuation for the robust $H_\infty$ estimation problem. We also have the augmented disturbance $\tilde{w}(t) := \left[\begin{array}{ccc}
w(t)^T & \frac{\epsilon_1}{\gamma}\eta(t)^T & \frac{\epsilon_2}{\gamma}\xi(t)^T
\end{array}\right]^T$, where $\eta(t) := F_1(t)Ex(t)$, and $\xi(t) := F_2(t)Gw(t)$.

\begin{theorem}\label{thm:h_infinity}
(See \cite{RP}) Consider the robust $H_\infty$ estimation problem for the uncertain system (\ref{eq:uncertain1}) converted to a scaled $H_\infty$ control problem for the system (\ref{eq:uncertain2}). Given a prescribed level of disturbance attenuation $\gamma > 0$, a robust $H_\infty$ estimator for the uncertain system (\ref{eq:uncertain1}) can be constructed, for some $\epsilon_1$, $\epsilon_2 > 0$, by solving the following two algebraic Riccati equations (AREs):

\vspace*{-3mm}\small
\begin{equation}\label{eq:robust_riccati1}
\begin{split}
\overline{A}^\dagger X&+X\overline{A}+X(\gamma^{-2} \overline{B}_1\overline{B}_1^\dagger)X
\\&+ \overline{C}_1^\dagger (I-\overline{D}_{12}\overline{E}_1^{-1}\overline{D}_{12}^\dagger)\overline{C}_1 = 0.
\end{split}
\end{equation}
\vspace*{-1mm}
\begin{equation}\label{eq:robust_riccati2}
\begin{split}
\overline{A}Y&+Y\overline{A}^\dagger +Y\overline{C}_1^\dagger\overline{C}_1Y+ \gamma^{-2}\overline{B}_1\overline{B}_1^\dagger\\
&-(\gamma^{-1}\overline{B}_1\overline{D}_{21}^\dagger+\gamma Y\overline{C}_2^\dagger)\\
&\times\overline{S}^\dagger\overline{E}_2^{-1}\overline{S}(\gamma^{-1}\overline{B}_1\overline{D}_{21}^\dagger+\gamma Y\overline{C}_2^\dagger)^\dagger = 0.
\end{split}
\end{equation}
\normalsize
where
\small
\begin{equation}\label{eq:final_parameters1}
\begin{split}
\overline{A} &= A, \, \overline{C}_2 = C, \, \overline{S} = S,\\
\overline{B}_1 &= \left[\begin{array}{ccc}
B(I-\epsilon_2^2G^\dagger G)^{-1/2} & \frac{\gamma}{\epsilon_1}H_1 & \frac{\gamma}{\epsilon_2}H_2
\end{array}\right],\\
\overline{C}_1 &= \left[\begin{array}{c}
\epsilon_1E\\
0\\
L
\end{array}\right], \, \overline{D}_{12} = \left[\begin{array}{c}
0\\
0\\
-I
\end{array}\right],\\
\overline{D}_{21} &= \left[\begin{array}{ccc}
D(I-\epsilon_2^2G^\dagger G)^{-1/2} & \frac{\gamma}{\epsilon_1}H_3 & 0
\end{array}\right],\\
\overline{E}_1 &= \overline{D}_{12}^\dagger\overline{D}_{12} = I,\\
\overline{E}_2 &= SD(I-\epsilon_2^2G^\dagger G)^{-1}D^\dagger S^\dagger + \frac{\gamma^2}{\epsilon_1^2}SH_3H_3^\dagger S^\dagger .
\end{split}
\end{equation}
\normalsize

A suitable estimator is then given by:
\vspace*{-2mm}\small
\begin{equation}\label{eq:robust_estimator}
\begin{split}
\dot{\hat{x}}(t) &= A_K\hat{x}(t) + B_Ky'(t),\\
\hat{z}(t) &= C_K\hat{x}(t),
\end{split}
\end{equation}
\normalsize
where
\vspace*{-2mm}\small
\begin{equation}\label{eq:rob_cls_est_matrices}
\begin{split}
A_K &= \overline{A} - B_K\overline{S}\overline{C}_2 + \gamma^{-2}(\overline{B}_1-B_K\overline{S}\overline{D}_{21})\overline{B}_1^\dagger X,\\
B_K &= \gamma^2(I-YX)^{-1}(Y\overline{C}_2^\dagger\overline{S}^\dagger + \gamma^{-2} \overline{B}_1\overline{D}_{21}^\dagger\overline{S}^\dagger)\overline{E}_2^{-1},\\
C_K &= -\overline{E}_1^{-1}\overline{D}_{12}^\dagger\overline{C}_1.
\end{split}
\end{equation}
\normalsize%\vspace*{1mm}
\end{theorem}

The estimation error is given as:

\vspace*{-1mm}\small
\begin{equation}
e(t) := \hat{z}(t)-z(t) = C_K\hat{x}(t)-Lx(t).
\end{equation}\normalsize

Then, the disturbance-to-error transfer function is \cite{RP}:
\vspace*{-1mm}\small
\begin{equation}\label{eq:error_spectrum}
\begin{split}
\tilde{G}_{we}(s) :=& \frac{e(s)}{w(s)} = \left[\begin{array}{cc}
-L & C_K
\end{array}\right]\\
\times&\left(sI - \left[\begin{array}{cc}
A+\Delta A(t) & 0\\
B_KS(C+\Delta C(t)) & A_K
\end{array}\right]\right)^{-1}\\
\times&\left[\begin{array}{c}
B+\Delta B(t)\\
B_KSD
\end{array}\right].
\end{split}
\end{equation}\normalsize

We are interested in the disturbance $\mathcal{A}$ to error $e$ transfer function, which is simply the first component of the matrix transfer function $\tilde{G}_{we}(s)$. We shall ignore the other component, which is the disturbance $\mathcal{A}^{\#}$ to error $e$ transfer function.

\section{Robust Coherent-Classical Estimation}
\begin{figure}[!t]
\vspace*{-3mm}
\centering
\includegraphics[width=0.45\textwidth]{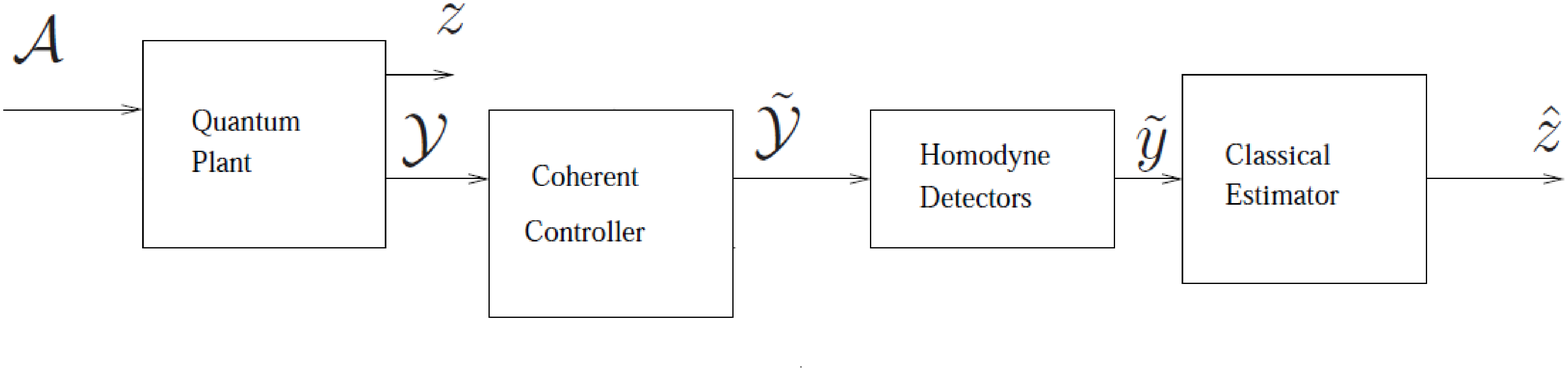}
\caption{Schematic diagram of coherent-classical estimation.}
\label{fig:coh_cls_scm}
\vspace*{-3mm}
\end{figure}

A schematic diagram of the coherent-classical estimation scheme is provided in Fig. \ref{fig:coh_cls_scm}. In this case, the plant output $\mathcal{Y}(t)$ does not directly drive a bank of homodyne detectors as in (\ref{eq:class_hd}). Rather, this output is fed into another quantum system called a coherent controller, defined as \cite{RPH}:

\vspace*{-3mm}\small
\begin{equation}\label{eq:coh_controller}
\begin{split}
\left[\begin{array}{c}
da_c(t)\\
da_c(t)^{\#}
\end{array}\right] &= A_c \left[\begin{array}{c}
a_c(t)\\
a_c(t)^{\#}
\end{array}\right] dt + B_c \left[\begin{array}{c}
d\mathcal{Y}(t)\\
d\mathcal{Y}(t)^{\#} 
\end{array}\right],\\
\left[\begin{array}{c}
d\tilde{\mathcal{Y}}(t)\\
d\tilde{\mathcal{Y}}(t)^{\#}
\end{array}\right] &= C_c \left[\begin{array}{c}
a_c(t)\\
a_c(t)^{\#}
\end{array}\right] dt + D_c \left[\begin{array}{c}
d\mathcal{Y}(t)\\
d\mathcal{Y}(t)^{\#} 
\end{array}\right].
\end{split}
\end{equation}
\normalsize

A quadrature of each component of $\tilde{\mathcal{Y}}(t)$ is homodyne detected to produce a corresponding classical signal $\tilde{y}_i$ \cite{RPH}:

\vspace*{-3mm}\small
\begin{equation}\label{eq:coh_class_hd}
\begin{split}
d\tilde{y}_1 &= \frac{e^{-\iota\tilde{\theta}_1}}{\sqrt{2}}d\tilde{\mathcal{Y}}_1 + \frac{e^{\iota\tilde{\theta}_1}}{\sqrt{2}}d\tilde{\mathcal{Y}}_1^{*},\\
&\vdots\\
d\tilde{y}_{\tilde{m}} &= \frac{e^{-\iota\tilde{\theta}_{\tilde{m}}}}{\sqrt{2}}d\tilde{\mathcal{Y}}_{\tilde{m}} + \frac{e^{\iota\tilde{\theta}_{\tilde{m}}}}{\sqrt{2}}d\tilde{\mathcal{Y}}_{\tilde{m}}^{*}.
\end{split}
\end{equation}\normalsize

Here, the angles $\tilde{\theta}_1,\hdots,\tilde{\theta}_{\tilde{m}}$ determine the quadrature measured by each homodyne detector. The vector of classical signals $\tilde{y} = [\tilde{y}_1 \hdots \tilde{y}_{\tilde{m}}]^T$ is then used as the input to a robust $H_\infty$ classical estimator of the form in the previous section.

The quantum plant (\ref{eq:plant}) augmented with the coherent controller (\ref{eq:coh_controller}) is defined by the QSDEs \cite{RPH}:

\vspace*{-3mm}\small
\begin{equation}\label{eq:coh_system1}
\begin{split}
\left[\begin{array}{c}
da(t)\\
da(t)^{\#}\\
da_c(t)\\
da_c(t)^{\#}
\end{array}\right] &= \left[\begin{array}{cc}
A & 0\\
B_cC & A_c
\end{array}\right] \left[\begin{array}{c}
a(t)\\
a(t)^{\#}\\
a_c(t)\\
a_c(t)^{\#}
\end{array}\right] dt\\
&+ \left[\begin{array}{c}
B\\
B_cD
\end{array}\right] \left[\begin{array}{c}
d\mathcal{A}(t)\\
d\mathcal{A}(t)^{\#}
\end{array}\right],\\
\left[\begin{array}{c}
d\tilde{\mathcal{Y}}(t)\\
d\tilde{\mathcal{Y}}(t)^{\#}
\end{array}\right] &= \left[\begin{array}{cc}
D_cC & C_c
\end{array}\right] \left[\begin{array}{c}
a(t)\\
a(t)^{\#}\\
a_c(t)\\
a_c(t)^{\#}
\end{array}\right] dt\\
&+ D_cD \left[\begin{array}{c}
d\mathcal{A}(t)\\
d\mathcal{A}(t)^{\#}
\end{array}\right].
\end{split}
\end{equation}
\normalsize\vspace*{-2mm}

The system (\ref{eq:coh_system1}) along with (\ref{eq:coh_class_hd}) is of the form:

\vspace*{-3mm}\small
\begin{equation}\label{eq:aug_system1}
\begin{split}
\dot{x}_a(t) = A_ax_a(t) + B_aw(t),\\
\tilde{y}'(t) = S_aC_ax_a(t) + S_aD_aw(t),
\end{split}
\end{equation}\normalsize
where $x_a(t) = \left[\begin{array}{cccc} a(t)^T & a(t)^{\dagger} & a_c(t)^T & a_c(t)^{\dagger}
\end{array}\right]^T$, $\tilde{y}'(t)$ is the measured output and

\vspace*{-3mm}\small
\begin{equation}\label{eq:aug_matrices1}
\begin{split}
A_a &= \left[\begin{array}{cc}
A & 0\\
B_cC & A_c
\end{array}\right], \, B_a = \left[\begin{array}{c}
B\\
B_cD
\end{array}\right],\\
C_a &= \left[\begin{array}{cc}
D_cC & C_c
\end{array}\right], \, D_a = D_cD, \, S_a = \left[\begin{array}{cc}
\tilde{S}_1 & \tilde{S}_2
\end{array}\right],\\
\tilde{S}_1 &= \left[\begin{array}{cccc}
\frac{e^{-\iota\tilde{\theta}_1}}{\sqrt{2}} & 0 & \hdots & 0\\
0 & \frac{e^{-\iota\tilde{\theta}_2}}{\sqrt{2}} & \hdots & 0\\
 & & \ddots & \\
 & & & \frac{e^{-\iota\tilde{\theta}_{\tilde{m}}}}{\sqrt{2}}
\end{array}\right],\\
\tilde{S}_2 &= \left[\begin{array}{cccc}
\frac{e^{\iota\tilde{\theta}_1}}{\sqrt{2}} & 0 & \hdots & 0\\
0 & \frac{e^{\iota\tilde{\theta}_2}}{\sqrt{2}} & \hdots & 0\\
 & & \ddots & \\
 & & & \frac{e^{\iota\tilde{\theta}_{\tilde{m}}}}{\sqrt{2}}
\end{array}\right].
\end{split}
\end{equation}
\normalsize\vspace*{-3mm}

Let us now consider an uncertain plant of the form (\ref{eq:uncertain1}). Then the $H_\infty$ estimation problem is for the following uncertain augmented system:

\vspace*{-3mm}\small
\begin{equation}
\begin{split}
\dot{x}_a(t) &= [A_a + \Delta A_a(t)] x_a(t) + [B_a + \Delta B_a(t)] w(t),\\
z(t) &= L_ax_a(t),\\
\tilde{y}'(t) &= S_a[C_a + \Delta C_a(t)] x_a(t) + S_aD_aw(t),
\end{split}
\end{equation}\normalsize
where we take $\Delta A_a(t) = H_{a1}F_1(t)E_a$, $\Delta B_a(t) = H_{a2}F_2(t)G_a$ and $\Delta C_a(t) = H_{a3}F_1(t)E_a$.

One can verify that these matrices can be expressed as:

\vspace*{-2mm}\small
\begin{equation}
\begin{split}
H_{a1} &= \left[\begin{array}{c}
H_1\\
B_cH_3
\end{array}\right], \, H_{a2} = \left[\begin{array}{c}
H_2\\
0
\end{array}\right], \, H_{a3} = D_cH_3,\\
E_a &= \left[\begin{array}{cc}
E & 0
\end{array}\right], \, G_a = G, \, L_a = \left[\begin{array}{cc}
L & 0
\end{array}\right].
\end{split}
\end{equation}\normalsize\vspace*{-2mm}

The robust $H_\infty$ estimator for the coherent-classical system is obtained by solving the following two AREs:

\vspace*{-3mm}\small
\begin{equation}\label{eq:aug_riccati1}
\begin{split}
\overline{A}_a^\dagger X_a&+X_a\overline{A}_a+X_a(\gamma^{-2} \overline{B}_{a1}\overline{B}_{a1}^\dagger)X_a\\
&+ \overline{C}_{a1}^\dagger (I-\overline{D}_{a12}\overline{E}_{a1}^{-1}\overline{D}_{a12}^\dagger)\overline{C}_{a1} = 0,
\end{split}
\end{equation}\vspace*{-1mm}
\begin{equation}\label{eq:aug_riccati2}
\begin{split}
\overline{A}_aY_a&+Y_a\overline{A}_a^\dagger +Y_a\overline{C}_{a1}^\dagger\overline{C}_{a1}Y_a +\gamma^{-2}\overline{B}_{a1}\overline{B}_{a1}^\dagger\\
&-(\gamma^{-1}\overline{B}_{a1}\overline{D}_{a21}^\dagger +\gamma Y_a\overline{C}_{a2}^\dagger)\\
&\times\overline{S}_a^\dagger\overline{E}_2^{-1}\overline{S}_a (\gamma^{-1}\overline{B}_{a1}\overline{D}_{a21}^\dagger +\gamma Y_a\overline{C}_{a2}^\dagger)^\dagger = 0,
\end{split}
\end{equation}\normalsize
which are of the forms (\ref{eq:robust_riccati1}) and (\ref{eq:robust_riccati2}), respectively.

Here, we have

\vspace*{-3mm}\small
\begin{equation}\label{eq:final_parameters5}
\begin{split}
\overline{A}_a &= A_a, \, \overline{C}_{a2} = C_a, \, \overline{S}_a = S_a,\\
\overline{B}_{a1} &= \left[\begin{array}{ccc}
B_a(I-\epsilon_2^2G_a^\dagger G_a)^{-1/2} & \frac{\gamma}{\epsilon_1}H_{a1} & \frac{\gamma}{\epsilon_2}H_{a2}
\end{array}\right],\\
\overline{C}_{a1} &= \left[\begin{array}{c}
\epsilon_1E_a\\
0\\
L_a
\end{array}\right],\, \overline{D}_{a12} = \left[\begin{array}{c}
0\\
0\\
-I
\end{array}\right],\\
\overline{D}_{a21} &= \left[\begin{array}{ccc}
D_a(I-\epsilon_2^2G_a^\dagger G_a)^{-1/2} & \frac{\gamma}{\epsilon_1}H_{a3} & 0
\end{array}\right].
\end{split}
\end{equation}\normalsize\vspace*{-2mm}

Then, a suitable robust estimator is given by
\vspace*{-2mm}\small
\begin{equation}\label{eq:robust_cohcls_estimator}
\begin{split}
\dot{\hat{x}}_a(t) &= A_{aK}\hat{x}_a(t) + B_{aK}\tilde{y}'(t),\\
\hat{z}(t) &= C_{aK}\hat{x}_a(t),
\end{split}
\end{equation}\normalsize
where
\vspace*{-1mm}\small
\begin{equation}\label{eq:robust_estimator_matrices}
\begin{split}
A_{aK} &= \overline{A}_a - B_{aK}\overline{S}_a\overline{C}_{a2} + \gamma^{-2}(\overline{B}_{a1}-B_{aK}\overline{S}_a\overline{D}_{a21})\overline{B}_{a1}^\dagger X_a,\\
B_{aK} &= \gamma^2(I-Y_aX_a)^{-1}(Y_a\overline{C}_{a2}^\dagger\overline{S}_a^\dagger + \gamma^{-2}\overline{B}_{a1}\overline{D}_{a21}^\dagger\overline{S}_a^\dagger) \overline{E}_{a2}^{-1},\\
C_{aK} &= -\overline{E}_{a1}^{-1}\overline{D}_{a12}^\dagger\overline{C}_{a1}.
\end{split}
\end{equation}
\normalsize\vspace*{-2mm}

Note that the matrices in (\ref{eq:final_parameters5}) of the robust coherent-classical estimator can be expressed in terms of the corresponding matrices in (\ref{eq:final_parameters1}) of the robust purely-classical estimator as follows:

\vspace*{-3mm}\small
\begin{equation}\label{eq:aug_mod_matrices1}
\begin{split}
\overline{A}_a &= \left[\begin{array}{cc}
\overline{A} & 0\\
B_c\overline{C}_2 & A_c
\end{array}\right], \, \overline{B}_{a1} = \left[\begin{array}{c}
\overline{B}_1\\
B_c\overline{D}_{21}
\end{array}\right],\\
\overline{C}_{a1} &= \left[\begin{array}{cc}
\overline{C}_1 & 0
\end{array}\right], \, \overline{C}_{a2} = \left[\begin{array}{cc}
D_c\overline{C}_2 & C_c
\end{array}\right],\\
\overline{D}_{a12} &= \overline{D}_{12}, \, \overline{D}_{a21} = D_c\overline{D}_{21}.
\end{split}
\end{equation}\normalsize

\section{Numerical Example}
We now present a numerical example. A linear quantum system arising in quantum optics is a dynamic squeezer - an optical cavity with a non-linear active medium inside. Let the plant be a dynamic squeezer, described by \cite{RPH}:

\vspace*{-3mm}\small
\begin{equation}\label{eq:sqz_plant}
\begin{split}
\left[\begin{array}{c}
da\\
da^{*}
\end{array}\right] &= \left[\begin{array}{cc}
-\frac{\beta}{2} & -\chi\\
-\chi^{*} & -\frac{\beta}{2}
\end{array}\right] \left[\begin{array}{c}
a\\
a^{*}
\end{array}\right] dt - \sqrt{\kappa} \left[\begin{array}{c}
d\mathcal{A}\\
d\mathcal{A}^{*}
\end{array}\right],\\
\left[\begin{array}{c}
d\mathcal{Y}\\
d\mathcal{Y}^{*}
\end{array}\right] &= \sqrt{\kappa} \left[\begin{array}{c}
a\\
a^{*}
\end{array}\right] dt + \left[\begin{array}{c}
d\mathcal{A}\\
d\mathcal{A}^{*}
\end{array}\right],\\
z(t) &= \left[\begin{array}{cc}
0.1 & -0.1
\end{array}\right] \left[\begin{array}{c}
a\\
a^{*}
\end{array}\right],
\end{split}
\end{equation}
\normalsize
where $\beta > 0$ is the overall cavity loss, $\kappa > 0$ determines the loss owing to the cavity mirrors, $\chi \in \mathbb{C}$ quantifies the non-linearity of the active medium, and $a$ is a single annihilation operator of the cavity mode.

Here, we choose $\beta = 4$, $\kappa = 4$, and $\chi = 0.5$. These values are chosen arbitrarily for the purposes of demonstration of principle here, and may well represent actual practical values for the corresponding physical parameters. We ensure though that the quantum system is physically realizable, since we have $\beta = \kappa$ (See \cite{IRP2,RPH,RPH1}). We fix the homodyne detection angle at $10^{\circ}$. Thus, the matrices in (\ref{eq:plant}) may be obtained.

We introduce uncertainty in the parameter $\alpha := \sqrt{\kappa}$ as follows: $\alpha \rightarrow \alpha+\mu\delta(t)\alpha$, where $|\delta(t)| \leq 1$ is an uncertain parameter and $\mu \in [0,1)$ is the level of uncertainty. Then,

\vspace*{-3mm}\small
\begin{equation}\label{eq:sqz_plant_unc1}
\begin{split}
\Delta A &= \left[\begin{array}{cc}
-\alpha^2\mu\delta-\frac{\alpha^2\mu^2\delta^2}{2} & 0\\
0 & -\alpha^2\mu\delta-\frac{\alpha^2\mu^2\delta^2}{2}
\end{array}\right],\\
\Delta B &= \left[\begin{array}{cc}
-\mu\delta\alpha & 0\\
0 & -\mu\delta\alpha
\end{array}\right], \, \Delta C = \left[\begin{array}{cc}
\mu\delta\alpha & 0\\
0 & \mu\delta\alpha
\end{array}\right].
\end{split}
\end{equation}
\normalsize

We define the relevant matrices as follows:

\vspace*{-3mm}\small
\begin{equation}\label{eq:sqz_plant_unc2}
\begin{split}
F_1(t) &= \left[\begin{array}{cccc}
\delta & 0 & 0 & 0\\
0 & \delta & 0 & 0\\
0 & 0 & \delta^2 & 0\\
0 & 0 & 0 & \delta^2
\end{array}\right], \, F_2(t) = \left[\begin{array}{cc}
\delta & 0\\
0 & \delta
\end{array}\right],\\
E &= \left[\begin{array}{cc}
-\frac{1}{2} & 0\\
0 & -\frac{1}{2}\\
-\frac{1}{2} & 0\\
0 & -\frac{1}{2}
\end{array}\right], \, G = \left[\begin{array}{cc}
1 & 0\\
0 & 1
\end{array}\right],\\
H_1 &= \left[\begin{array}{cccc}
2\mu\alpha^2 & 0 & \mu^2\alpha^2 & 0\\
0 & 2\mu\alpha^2 & 0 & \mu^2\alpha^2
\end{array}\right],\\
H_2 &= \left[\begin{array}{cc}
-\mu\alpha & 0\\
0 & -\mu\alpha
\end{array}\right],\\
H_3 &= \left[\begin{array}{cccc}
-2\mu\alpha & 0 & 0 & 0\\
0 & -2\mu\alpha & 0 & 0
\end{array}\right].
\end{split}
\end{equation}
\normalsize\vspace*{-1mm}

One can verify we have $\Delta A(t) = H_1F_1(t)E$, $\Delta B(t) = H_2F_2(t)G$ and $\Delta C(t) = H_3F_1(t)E$, as required in (\ref{eq:unc_pars}). We fix $\delta = -1$, such that (\ref{eq:unc_constraint}) is satisfied, and set $\mu = 0.1$.

We now solve the associated robust purely-classical $H_\infty$ estimation problem using Theorem \ref{thm:h_infinity}. We set $\gamma = 0.65$ and choose $\epsilon_1 = 0.19$, $\epsilon_2 = 0.81$, such that $\eta(t)$ and $\xi(t)$ are sufficiently suppressed, yielding satisfactory solutions to (\ref{eq:robust_riccati1}), (\ref{eq:robust_riccati2}). Then, an estimator is obtained as in (\ref{eq:robust_estimator}) with:

\vspace*{-3mm}\small
\begin{equation}
\begin{split}
A_K &= \left[\begin{array}{cc}
-0.0274-2.3799\iota & 1.8584-1.6718\iota\\
1.8584+1.6718\iota & -0.0274+2.3799\iota
\end{array}\right],\\
B_K &= \left[\begin{array}{c}
-1.5600+1.5188\iota\\
-1.5600-1.5188\iota
\end{array}\right], \, C_K = \left[\begin{array}{cc}
0.1 & -0.1
\end{array}\right].
\end{split}
\end{equation}
\normalsize\vspace*{-3mm}

Now, let the controller be another dynamic squeezer \cite{RPH}:

\vspace*{-3mm}\small
\begin{equation}\label{eq:sqz_coh_controller}
\begin{split}
\left[\begin{array}{c}
da_c\\
da_c^{*}
\end{array}\right] &= \left[\begin{array}{cc}
-\frac{\beta_c}{2} & -\chi_c\\
-\chi_c^{*} & -\frac{\beta_c}{2}
\end{array}\right] \left[\begin{array}{c}
a_c\\
a_c^{*}
\end{array}\right] dt - \sqrt{\kappa_c} \left[\begin{array}{c}
d\mathcal{Y}\\
d\mathcal{Y}^{*}
\end{array}\right],\\
\left[\begin{array}{c}
d\tilde{\mathcal{Y}}\\
d\tilde{\mathcal{Y}}^{*}
\end{array}\right] &= \sqrt{\kappa_c} \left[\begin{array}{c}
a_c\\
a_c^{*}
\end{array}\right] dt + \left[\begin{array}{c}
d\mathcal{Y}\\
d\mathcal{Y}^{*}
\end{array}\right],
\end{split}
\end{equation}
\normalsize%\vspace*{-3mm}
where we choose $\beta_c = 4$, $\kappa_c = 4$, $\chi_c = -1$, such that it is physically realizable, and then obtain the matrices in (\ref{eq:coh_controller}).

We now consider the plant to be uncertain as in (\ref{eq:sqz_plant_unc1}), (\ref{eq:sqz_plant_unc2}). Also, we fix $\delta=-1$, $\mu = 0.1$ and the homodyne detection angle at $10^{\circ}$. We again set $\gamma = 0.65$, and choose $\epsilon_1 = 0.19$, $\epsilon_2 = 0.81$. A robust coherent-classical estimator is then obtained as in (\ref{eq:robust_cohcls_estimator}) with:

\vspace*{-3mm}\tiny
\begin{equation}
\begin{split}
A_{aK} &= \left[\begin{array}{cccc}
-2.03 + 0.13\iota & -0.86 + 0.03\iota & -0.28 + 0.14\iota & -0.31 + 0.03\iota\\
-0.86 - 0.03\iota & -2.03 - 0.13\iota & -0.31 - 0.03\iota & -0.28 - 0.14\iota\\
-6.59 + 0.50\iota & -2.90 - 0.47\iota & -4.95 + 0.54\iota & -1.95 - 0.50\iota\\
-2.90 + 0.47\iota & -6.59 - 0.50\iota & -1.95 + 0.50\iota & -4.95 - 0.54\iota
\end{array}\right],\\
B_{aK} &= \left[\begin{array}{c}
0.21 - 0.06\iota\\
0.21 + 0.06\iota\\
2.12 - 0.01\iota\\
2.12 + 0.01\iota
\end{array}\right], \, C_{aK} = \left[\begin{array}{cccc}
0.1 & -0.1 & 0 & 0
\end{array}\right].
\end{split}
\end{equation}
\normalsize\vspace*{-3mm}

\begin{figure}[!t]
\vspace*{-3mm}
%\centering
\includegraphics[width=0.45\textwidth]{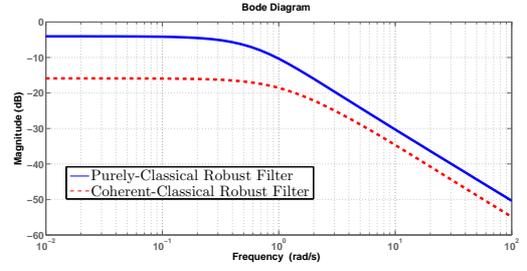}
\caption{Comparison of the estimation error frequency response of robust coherent-classical and robust purely-classical estimators.}
\label{fig:errors_spectra1}
%\vspace*{-3mm}
\end{figure}

\begin{figure}[!t]
\vspace*{-3mm}
%\centering
\includegraphics[width=0.45\textwidth]{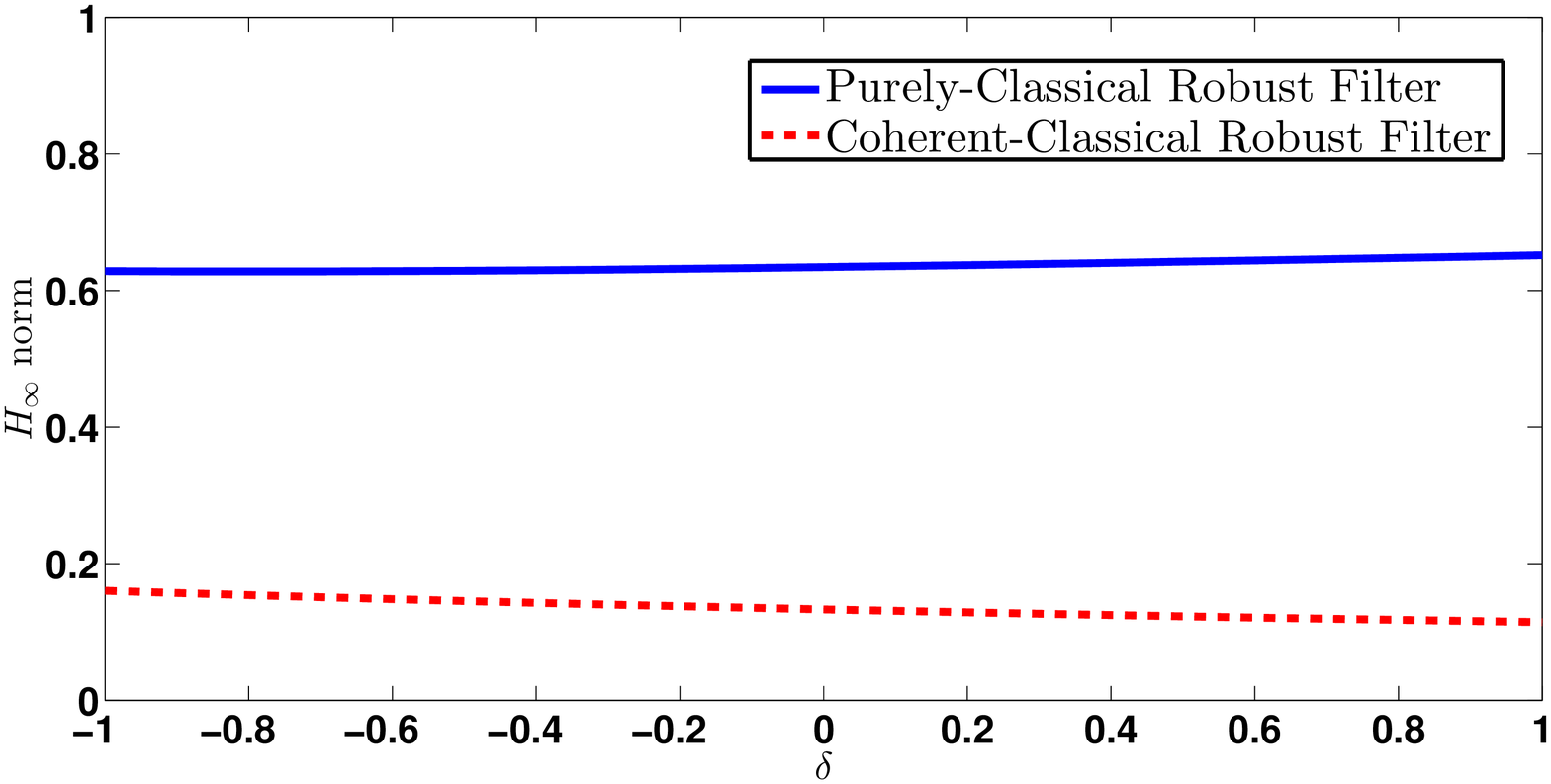}
\caption{Comparison of the $H_\infty$ norm of the disturbance-to-error transfer functions for robust coherent-classical and robust purely-classical estimators.}
\label{fig:errors_hnorm1}
\vspace*{-3mm}
\end{figure}

Fig. \ref{fig:errors_spectra1} shows a comparison of the error spectra (bode magnitude plots of the disturbance $\mathcal{A}$ to error $e$ transfer function only from (\ref{eq:error_spectrum}) for augmented plant-controller system and plant alone respectively) of the robust coherent-classical filter and the robust purely-classical filter. Clearly, the robust $H_\infty$ coherent-classical filter provides better disturbance attenuation compared to the robust $H_\infty$ purely-classical filter. Fig. \ref{fig:errors_hnorm1} shows a comparison of the $H_\infty$ norm of the disturbance-to-error transfer functions as a function of $\delta \in [-1,1]$ for the two robust filters. Clearly, the robust coherent-classical filter provides higher disturbance attenuation than the robust classical-only filter across the entire uncertainty window.


\begin{figure}[!b]
\vspace*{-5mm}
\centering
\includegraphics[width=0.45\textwidth]{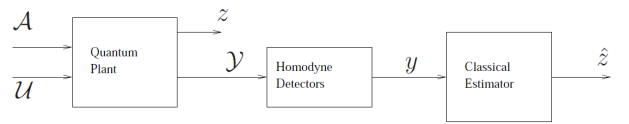}
\caption{Modified schematic diagram of purely-classical estimation.}
\label{fig:fb_cls_scm}
\vspace*{-1mm}
\end{figure}

\section{Coherent Feedback Case}
Here, we consider the case where there is quantum feedback from the controller to the plant \cite{IRP2,RPH1}. For this purpose, the plant is assumed to have an additional control input $\mathcal{U}$ (See Fig. \ref{fig:fb_cls_scm}). The plant (\ref{eq:plant}) then is of the form \cite{RPH1}:

\vspace*{-3mm}\small
\begin{equation}\label{eq:fb_plant}
\begin{split}
\left[\begin{array}{c}
da(t)\\
da(t)^{\#}
\end{array}\right] &= A \left[\begin{array}{c}
a(t)\\
a(t)^{\#}
\end{array}\right] dt + \left[\begin{array}{cc}
B_1 & B_2
\end{array}\right] \left[\begin{array}{c}
d\mathcal{A}(t)\\
d\mathcal{A}(t)^{\#}\\
d\mathcal{U}(t)\\
d\mathcal{U}(t)^{\#}
\end{array}\right],\\
\left[\begin{array}{c}
d\mathcal{Y}(t)\\
d\mathcal{Y}(t)^{\#}
\end{array}\right] &= C \left[\begin{array}{c}
a(t)\\
a(t)^{\#}
\end{array}\right] dt + \left[\begin{array}{cc}
D & 0
\end{array}\right] \left[\begin{array}{c}
d\mathcal{A}(t)\\
d\mathcal{A}(t)^{\#}\\
d\mathcal{U}(t)\\
d\mathcal{U}(t)^{\#}
\end{array}\right],\\
z &= L\left[\begin{array}{c}
a(t)\\
a(t)^{\#} 
\end{array}\right].
\end{split}
\end{equation}
\normalsize\vspace*{-3mm}

\begin{figure}%[!b]
%\vspace*{-1mm}
\centering
\includegraphics[width=0.45\textwidth]{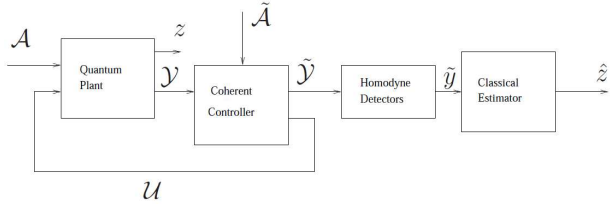}
\caption{Schematic of coherent-classical estimation with coherent feedback.}
\label{fig:fb_coh_cls_scm}
\vspace*{-5mm}
\end{figure}

The uncertain plant along with (\ref{eq:class_hd}) is then modelled as:

\vspace*{-3mm}\small
\begin{equation}\label{eq:uncertain_fb1}
\begin{split}
\dot{x}(t) &= [A+\Delta A(t)]x(t) + \left[\begin{array}{cc}
B_1+\Delta B_1(t) & B_2
\end{array}\right]\overline{w}(t),\\
z(t) &= Lx(t),\\
y'(t) &= S[C+\Delta C(t)]x(t) + S\left[\begin{array}{cc}
D & 0
\end{array}\right]\overline{w}(t),
\end{split}
\end{equation}
\normalsize
where $\overline{w}(t) := \left[\begin{array}{cc}
w(t)^T & u_c(t)^T
\end{array}\right]^T$, and $u_c(t)$ is the spare control input. Also, we have $\Delta A(t) = H_1F_1(t)E$, $\Delta B_1(t) = H_2F_2(t)G$, $\Delta C(t) = H_3F_1(t)E$. The robust purely-classical estimator is then obtained from Theorem \ref{thm:h_infinity}, where we have:

\vspace*{-3mm}\small
\begin{equation}\label{eq:coh_cls_thm_matrices}
\begin{split}
\overline{A} &= A, \, \overline{C}_2 = C, \, \overline{S} = S,\\
\overline{B}_1 &= \left[\begin{array}{cccc}
B_1(I-\epsilon_2^2G^\dagger G)^{-1/2} & B_2 & \frac{\gamma}{\epsilon_1}H_1 & \frac{\gamma}{\epsilon_2}H_2
\end{array}\right],\\
\overline{C}_1 &= \left[\begin{array}{c}
\epsilon_1E\\
0\\
L
\end{array}\right], \, \overline{D}_{12} = \left[\begin{array}{c}
0\\
0\\
-I
\end{array}\right],\\
\overline{D}_{21} &= \left[\begin{array}{cccc}
D(I-\epsilon_2^2G^\dagger G)^{-1/2} & 0 & \frac{\gamma}{\epsilon_1}H_3 & 0
\end{array}\right].
\end{split}
\end{equation}
\normalsize

The controller here would have an additional output that is fed back to the control input of the plant (See Fig. \ref{fig:fb_coh_cls_scm}). The controller is defined as \cite{IRP2,RPH1}:

\vspace*{-3mm}\small
\begin{equation}\label{eq:fb_coh_controller}
\begin{split}
\left[\begin{array}{c}
da_c(t)\\
da_c(t)^{\#}
\end{array}\right] &= A_c \left[\begin{array}{c}
a_c(t)\\
a_c(t)^{\#}
\end{array}\right] dt\\
&+ \left[\begin{array}{cc}
B_{c1} & B_{c2}
\end{array}\right] \left[\begin{array}{c}
d\tilde{\mathcal{A}}(t)\\
d\tilde{\mathcal{A}}(t)^{\#}\\
d\mathcal{Y}(t)\\
d\mathcal{Y}(t)^{\#} 
\end{array}\right],\\
\left[\begin{array}{c}
d\tilde{\mathcal{Y}}(t)\\
d\tilde{\mathcal{Y}}(t)^{\#}\\
d\mathcal{U}(t)\\
d\mathcal{U}(t)^{\#}
\end{array}\right] &= \left[\begin{array}{c}
\tilde{C}_c\\
C_c
\end{array}\right] \left[\begin{array}{c}
a_c(t)\\
a_c(t)^{\#}
\end{array}\right] dt\\
&+ \left[\begin{array}{cc}
\tilde{D}_{c1} & \tilde{D}_{c2}\\
D_{c1} & D_{c2}
\end{array}\right] \left[\begin{array}{c}
d\tilde{\mathcal{A}}(t)\\
d\tilde{\mathcal{A}}(t)^{\#}\\
d\mathcal{Y}(t)\\
d\mathcal{Y}(t)^{\#} 
\end{array}\right].
\end{split}
\end{equation}
\normalsize\vspace*{-3mm}

The plant (\ref{eq:fb_plant}) and the controller (\ref{eq:fb_coh_controller}) can be combined to yield an augmented system \cite{IRP2,RPH1}:

\vspace*{-3mm}\small
\begin{equation}\label{eq:fb_coh_system}
\begin{split}
\left[\begin{array}{c}
da(t)\\
da(t)^{\#}\\
da_c(t)\\
da_c(t)^{\#}
\end{array}\right] &= \left[\begin{array}{cc}
A+B_2D_{c2}C & B_2C_c\\
B_{c2}C & A_c
\end{array}\right] \left[\begin{array}{c}
a(t)\\
a(t)^{\#}\\
a_c(t)\\
a_c(t)^{\#}
\end{array}\right] dt\\
&+ \left[\begin{array}{cc}
B_1+B_2D_{c2}D & B_2D_{c1}\\
B_{c2}D & B_{c1}
\end{array}\right] \left[\begin{array}{c}
d\mathcal{A}(t)\\
d\mathcal{A}(t)^{\#}\\
d\tilde{\mathcal{A}}(t)\\
d\tilde{\mathcal{A}}(t)^{\#}
\end{array}\right],\\
\left[\begin{array}{c}
d\tilde{\mathcal{Y}}(t)\\
d\tilde{\mathcal{Y}}(t)^{\#}
\end{array}\right] &= \left[\begin{array}{cc}
\tilde{D}_{c2}C & \tilde{C}_c
\end{array}\right] \left[\begin{array}{c}
a(t)\\
a(t)^{\#}\\
a_c(t)\\
a_c(t)^{\#}
\end{array}\right] dt\\
&+ \left[\begin{array}{cc}
\tilde{D}_{c2}D & \tilde{D}_{c1}
\end{array}\right] \left[\begin{array}{c}
d\mathcal{A}(t)\\
d\mathcal{A}(t)^{\#}\\
d\tilde{\mathcal{A}}(t)\\
d\tilde{\mathcal{A}}(t)^{\#}
\end{array}\right].
\end{split}
\end{equation}
\normalsize\vspace*{-3mm}

The augmented system (\ref{eq:fb_coh_system}) along with (\ref{eq:coh_class_hd}) is of the form (\ref{eq:aug_system1}), (\ref{eq:aug_matrices1}), with $w(t)$ replaced by $w'(t) := \left[\begin{array}{cccc}
d\mathcal{A}(t)^T & d\mathcal{A}(t)^{\dagger} & d\tilde{\mathcal{A}}(t)^T & d\tilde{\mathcal{A}}(t)^{\dagger}
\end{array}\right]^T$, and with

\vspace*{-3mm}\small
\begin{equation}
\begin{split}
A_a &= \left[\begin{array}{cc}
A+B_2D_{c2}C & B_2C_c\\
B_{c2}C & A_c
\end{array}\right],\\
B_a &= \left[\begin{array}{cc}
B_1+B_2D_{c2}D & B_2D_{c1}\\
B_{c2}D & B_{c1}
\end{array}\right],\\
C_a &= \left[\begin{array}{cc}
\tilde{D}_{c2}C & \tilde{C}_c
\end{array}\right], \, D_a = \left[\begin{array}{cc}
\tilde{D}_{c2}D & \tilde{D}_{c1}
\end{array}\right].
\end{split}
\end{equation}
\normalsize

Let us now consider an uncertain plant of the form (\ref{eq:uncertain_fb1}). Then the $H_\infty$ estimation problem is considered for the following uncertain augmented system:

\vspace*{-3mm}\small
\begin{equation}
\begin{split}
\dot{x}_a(t) &= [A_a + \Delta A_a(t)] x_a(t) + [B_a + \Delta B_a(t)] w'(t),\\
z(t) &= L_ax_a(t),\\
\tilde{y}'(t) &= S_a[C_a + \Delta C_a(t)] x_a(t) + S_aD_aw'(t),
\end{split}
\end{equation}
\normalsize
where we take $\Delta A_a(t) = H_{a1}F_1(t)E_a$, $\Delta B_a(t) = H_{a2}F_2(t)G_a$ and $\Delta C_a(t) = H_{a3}F_1(t)E_a$. One can verify that these matrices can be expressed as follows:

\vspace*{-3mm}\small
\begin{equation}
\begin{split}
H_{a1} &= \left[\begin{array}{c}
H_1+B_2D_{c2}H_3\\
B_{c2}H_3
\end{array}\right], \, H_{a2} = \left[\begin{array}{c}
H_2\\
0
\end{array}\right], \, H_{a3} = \tilde{D}_{c2}H_3,\\
E_a &= \left[\begin{array}{cc}
E & 0
\end{array}\right], \, G_a = \left[\begin{array}{cc}
G & 0
\end{array}\right], \, L_a = \left[\begin{array}{cc}
L & 0
\end{array}\right].
\end{split}
\end{equation}
\normalsize

The robust $H_\infty$ estimator for the coherent-classical system here is then obtained by solving two AREs of the form (\ref{eq:aug_riccati1}) and (\ref{eq:aug_riccati2}) with the relevant matrices as defined in (\ref{eq:final_parameters5}). A suitable estimator is as defined in (\ref{eq:robust_cohcls_estimator}), (\ref{eq:robust_estimator_matrices}).

%\section{Numerical Example 2}
Consider an example with the plant given by \cite{IRP2,RPH1}:

\vspace*{-3mm}\small
\begin{equation}\label{eq:fb_sqz_plant}
\begin{split}
\left[\begin{array}{c}
da\\
da^{*}
\end{array}\right] &= \left[\begin{array}{cc}
-\frac{\beta}{2} & -\chi\\
-\chi^{*} & -\frac{\beta}{2}
\end{array}\right] \left[\begin{array}{c}
a\\
a^{*}
\end{array}\right] dt\\
&- \sqrt{\kappa_1} \left[\begin{array}{c}
d\mathcal{A}\\
d\mathcal{A}^{*}
\end{array}\right] - \sqrt{\kappa_2} \left[\begin{array}{c}
d\mathcal{U}\\
d\mathcal{U}^{*}
\end{array}\right],\\
\left[\begin{array}{c}
d\mathcal{Y}\\
d\mathcal{Y}^{*}
\end{array}\right] &= \sqrt{\kappa_1} \left[\begin{array}{c}
a\\
a^{*}
\end{array}\right] dt + \left[\begin{array}{c}
d\mathcal{A}\\
d\mathcal{A}^{*}
\end{array}\right],\\
z(t) &= \left[\begin{array}{cc}
0.1 & -0.1
\end{array}\right] \left[\begin{array}{c}
a\\
a^{*}
\end{array}\right].
\end{split}
\end{equation}
\normalsize\vspace*{-3mm}

Here, we choose $\beta = 4$, $\kappa_1 = \kappa_2 = 2$ and $\chi = -1$. Note that this system is physically realizable, since $\beta = \kappa_1+\kappa_2$. The matrices in (\ref{eq:fb_plant}) may then be obtained.

The coherent controller (\ref{eq:sqz_coh_controller}) is of the form \cite{IRP2,RPH1}:

\vspace*{-3mm}\small
\begin{equation}\label{eq:fb_sqz_ctrlr}
\begin{split}
\left[\begin{array}{c}
da_c\\
da_c^{*}
\end{array}\right] &= \left[\begin{array}{cc}
-\frac{\beta_c}{2} & -\chi_c\\
-\chi_c^{*} & -\frac{\beta_c}{2}
\end{array}\right] \left[\begin{array}{c}
a_c\\
a_c^{*}
\end{array}\right] dt\\
&- \sqrt{\kappa_{c1}} \left[\begin{array}{c}
d\tilde{\mathcal{A}}\\
d\tilde{\mathcal{A}}^{*}
\end{array}\right] - \sqrt{\kappa_{c2}} \left[\begin{array}{c}
d\mathcal{Y}\\
d\mathcal{Y}^{*}
\end{array}\right],\\
\left[\begin{array}{c}
d\tilde{\mathcal{Y}}\\
d\tilde{\mathcal{Y}}^{*}
\end{array}\right] &= \sqrt{\kappa_{c1}} \left[\begin{array}{c}
a_c\\
a_c^{*}
\end{array}\right] dt + \left[\begin{array}{c}
d\tilde{\mathcal{A}}\\
d\tilde{\mathcal{A}}^{*}
\end{array}\right],\\
\left[\begin{array}{c}
d\mathcal{U}\\
d\mathcal{U}^{*}
\end{array}\right] &= \sqrt{\kappa_{c2}} \left[\begin{array}{c}
a_c\\
a_c^{*}
\end{array}\right] dt + \left[\begin{array}{c}
d\mathcal{Y}\\
d\mathcal{Y}^{*}
\end{array}\right].
\end{split}
\end{equation}
\normalsize\vspace*{-3mm}

Here, we choose $\beta_c = 4$, $\kappa_{c1} = \kappa_{c2} = 2$ and $\chi_c = 0.5$. Note that this system is physically realizable since $\beta_c = \kappa_{c1} + \kappa_{c2}$. Then the matrices in (\ref{eq:fb_coh_controller}) may be obtained.

We now consider the plant to be uncertain as in (\ref{eq:uncertain_fb1}), (\ref{eq:sqz_plant_unc2}) with $\alpha = \sqrt{\kappa_1}$ here. We set $\delta = -1$, $\mu = 0.1$ and fix the homodyne detection angle at $80^{\circ}$. Also, we choose $\gamma = 0.65$, $\epsilon_1 = 0.2$, $\epsilon_2 = 0.6$. A robust $H_\infty$ purely-classical estimator is obtained as in (\ref{eq:robust_estimator}), (\ref{eq:coh_cls_thm_matrices}) with:

\vspace*{-3mm}\small
\begin{equation}
\begin{split}
A_{K} &= \left[\begin{array}{cc}
1.4589 + 1.4235\iota & -2.5630 - 0.2493\iota\\
-2.5630 + 0.2493\iota & 1.4589 - 1.4235\iota
\end{array}\right],\\
B_{K} &= \left[\begin{array}{c}
0.8659 - 3.6066\iota\\
0.8659 + 3.6066\iota
\end{array}\right], \, C_{K} = \left[\begin{array}{cc}
0.1 & -0.1
\end{array}\right].
\end{split}
\end{equation}
\normalsize\vspace*{-3mm}

Also, a robust $H_\infty$ coherent-classical estimator is obtained as in (\ref{eq:robust_cohcls_estimator}) with:

\vspace*{-3mm}\tiny
\begin{equation}
\begin{split}
A_{aK} &= \left[\begin{array}{cccc}
-3.88 - 0.002\iota & 1.05 - 0.003\iota & -4.29 - 0.27\iota & 2.17 - 0.51\iota\\
1.05 + 0.003\iota & -3.88 + 0.002\iota & 2.17 + 0.51\iota & -4.29 + 0.27\iota\\
-1.95 - 0.003\iota & 0.03 - 0.004\iota & -4.98 - 0.32\iota & 2.39 - 0.72\iota\\
0.03 + 0.004\iota & -1.95 + 0.003\iota & 2.39 + 0.72\iota & -4.98 + 0.32\iota
\end{array}\right],\\
B_{aK} &= \left[\begin{array}{c}
0.12 + 2.267\iota\\
0.12 - 2.267\iota\\
0.20 + 2.993\iota\\
0.20 - 2.993\iota
\end{array}\right], \, C_{aK} = \left[\begin{array}{cccc}
0.1 & -0.1 & 0 & 0
\end{array}\right].
\end{split}
\end{equation}
\normalsize\vspace*{-3mm}

\begin{figure}[!t]
\vspace*{-3mm}
%\centering
\includegraphics[width=0.45\textwidth]{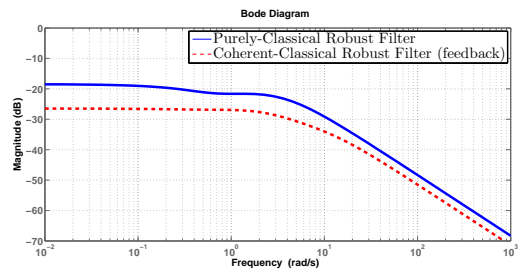}
\caption{Comparison of the estimation error frequency response of robust coherent-classical and robust purely-classical estimators.}
\label{fig:errors_spectra2}
%\vspace*{-3mm}
\end{figure}

\begin{figure}[!t]
\vspace*{-3mm}
%\centering
\includegraphics[width=0.45\textwidth]{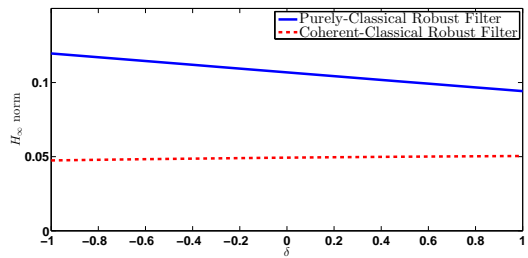}
\caption{Comparison of the $H_\infty$ norm of the disturbance-to-error transfer functions for robust coherent-classical and robust purely-classical estimators.}
\label{fig:errors_hnorm2}
\vspace*{-3mm}
\end{figure}

Fig. \ref{fig:errors_spectra2} illustrates that the robust $H_\infty$ coherent-classical filter provides with better disturbance attenuation compared to the robust $H_\infty$ purely-classical filter for the uncertain plant. Fig. \ref{fig:errors_hnorm2} shows that this holds for all values of $\delta$. Moreover, we can see that the $H_\infty$ norm for the robust coherent-classical estimator is uniform across the uncertainty window unlike the robust purely-classical estimator. That is, our robust coherent-classical estimator exhibits improved robustness to uncertainty as compared to the robust classical-only estimator. This is due to the coherent feedback involved here as opposed to the case of Fig. \ref{fig:errors_hnorm1}.

Note that we here considered uncertainty in $B_1$ only in (\ref{eq:uncertain_fb1}). This is because any uncertainty in $B_1$ (and not $B_2$) in (\ref{eq:fb_plant}) will also cause $C$ (besides $A$) to be accordingly uncertain owing to physical realizability constraints for our example (\ref{eq:fb_sqz_plant}). However, uncertainty in $B_2$ only and/or uncertainties in both $B_1$ and $B_2$ can be treated as well.


\section{Conclusion}
In this paper, we studied robust $H_\infty$ coherent-classical estimation, with and without coherent feedback, and compared with robust $H_\infty$ purely-classical estimation for an uncertain linear quantum plant. We observed that our robust coherent-classical filter, whether or not involving coherent feedback, can provide better disturbance attenuation than the purely-classical filter. Additionally, with coherent feedback, our robust coherent-classical filter provides superior robustness to uncertainty compared to the classical-only filter.

%\section*{Acknowledgment}
%The first author would like to thank Dr. Mohamed Mabrok for useful technical discussion with regards to this work.

\bibliographystyle{IEEEtran}
\bibliography{IEEEabrv,aucc16bib}

\end{document}